\documentstyle[11pt,epsfig,amssymb,multirow]{article}
\newcommand{\sta}{{\rm sta}}

\newcommand{\vsig}{\varsigma}

\newcommand{\qq}{q}

\newcommand{\real}{{\mathbb R}} 

\newcommand{\half}{\frac{1}{2}}

\newcommand{\bvsig}{\mbox{\boldmath$\varsigma$}}

\newcommand{\bveps}{\mbox{\boldmath$\xi$}}

\newcommand{\barbvsig}{\bar{\mbox{\boldmath$\varsigma$}}}

\newcommand{\btau}{{\mbox{\boldmath$\tau$}}}

\newcommand{\barbtau}{\bar{\mbox{\boldmath$\tau$}}}
\newcommand{\bchi}{{\mbox{\boldmath$\chi$}}}

\newcommand{\barbchi}{\bar{\mbox{\boldmath$\chi$}}}
\newcommand{\bxi}{{\mbox{\boldmath$\xi$}}}

\newcommand{\Oo}{\Omega}

\newcommand{\Gt}{{\Gamma_t}}

\newcommand{\bE}{{\bf E}}
\newcommand{\bQ}{{\bf Q}}

\newcommand{\eba}{\begin{array}}
\newcommand{\eea}{\end{array}}
\newcommand{\ebe}{\begin{eqnarray}}
\newcommand{\eee}{\end{eqnarray}}
\newcommand{\eb}{\begin{equation}}
\newcommand{\ee}{\end{equation}}
\newcommand{\bT}{{\bf T}}

\newcommand{\calP}{{\cal{P}}}

\newcommand{\bC}{{\bf C}}

\newcommand{\bH}{{\bf H}}
\newcommand{\bn}{{\bf n}}
\newcommand{\bt}{{\bf t}}
\newcommand{\bff}{{\bf f}}

\newcommand{\bX}{{\bf X}}

\newcommand{\bF}{{\bf F}}

\newcommand{\bB}{{\bf B}}
\newcommand{\bI}{{\bf I}}

\newcommand{\calS}{{\cal S}}

\newcommand{\calF}{{\cal F}}

\newcommand{\calL}{{\cal L}}

\newcommand{\calT}{{\cal T}}

\newcommand{\calE}{{\cal E}}

\newcommand{\calX}{{\cal X}}

\newcommand{\barbS}{\bar{\bf S}}
\newcommand{\barbT}{\bar{\bf T}}

\newcommand{\bS}{{\bf S}}

\newcommand{\dO}{\,\mbox{d}\Oo}
\newcommand{\dG}{\,\mbox{d} \Gamma}

\newcommand{ \grad }{{\mbox{grad}}}
\newcommand{\alp}{{\alpha}}

\newcommand{ \sig}{{\sigma}}
\newcommand{ \Lam}{{\Lambda}}

\newcommand{ \lam}{{\lambda}}

\newcommand{\tr}{{\mbox{tr}}}

\newtheorem{thm}{Theorem}[section]
\newtheorem{definition}{Definition}[section]

    \newcommand{\bLam}{{\bf \Lambda}}
   \newcommand{\UU}{U}
    
   \newcommand{\WW}{W}

 \newcommand{\SO}{\mbox{SO}}

\newtheorem{rem}{Remark}[section]
\newtheorem{defi}{Definition}[section]
\newtheorem{cor}{Corollary}[section]

\newtheorem{lem}{Lemma}[section]

    \renewcommand{\bT}{{\bf S}}
    \renewcommand{\barbT}{\bar{\bf S}}

\begin{document}
$\;$\\
{\Large \textbf{Analytic Solutions to 3-D Finite Deformation Problems\\ Governed by  St Venant-Kirchhoff Material
}}\\
\\
{\textbf{  David Yang Gao$^{1,2}$  $\&$  Eldar Hajilarov$^1$
}}\\
{\em 1.  Faculty of Science and Technology,Federation University Australia,  
Mt Helen,  VIC 3353, Australia.\\
2. Research School of Engineering, Australian National University, Canberra, Australia}

\begin{abstract}

This paper presents a detailed study  on analytical solutions to
a general nonlinear boundary-value problem  in finite deformation theory.
  Based on  canonical duality theory and the associated pure complementary energy principle in nonlinear elasticity
  proposed by Gao in 1999,
   we  show that the general  nonlinear partial differential equation for deformation   is actually equivalent to
   an   algebraic (tensor) equation in stress space. For   St Venant-Kirchhoff
materials, this coupled cubic algebraic equation can be solved principally
     to obtain all possible solutions. Our results show that for any given   external source field
     such that the statically admissible first Piola-Kirchhoff stress field has no-zero eigenvalues,
     the problem has  a unique   global minimal solution, which is
      corresponding to a positive-definite second Piola-Kirchhoff stress $\bT$,   and
     at most eight local solutions corresponding to negative-definite $\bT$. Additionally, the problem could have 15 unstable      solutions corresponding to indefinite $\bT$.
 This paper demonstrates that the canonical duality theory and the pure complementary energy principle   play
fundamental roles in nonconvex analysis and finite deformation theory.
\end{abstract}


\section{Nonconvex Variational Problem and Motivation}

A large class of  finite deformation problems in nonlinear elasticity can be
formulated on the basis of a variational principle $(\calP)$ in
which it is required to minimize certain nonconvex  potential energy.  Typically,
this takes the form
\begin{equation}
(\calP): \;\; \min_{\bchi \in \calX_a} \left\{ \Pi(\bchi) = \int_{\Omega}
W(\nabla \bchi) {\rm d} \Omega + \int_{\Omega} \phi(\bchi) \rho
{\rm d} \Omega - \int_{\Gamma_t} \bchi \cdot \bt {\rm d} \Gamma
\right\} , \label{pprobm}
\end{equation}
where $\bchi$ represents the deformation field (a bijection),
$W(\bF)$ is the strain energy per unit reference volume, which is
a nonlinear differentiable function of the deformation gradient
$\bF=\nabla \bchi$,  and $\nabla$ is the gradient operator in
a simply-connected domain (the reference configuration of the body)
$\Omega \subset \real^3  $
with boundary $\partial \Omega = \Gamma = \Gamma_t \cup
\Gamma_\chi$ such that $\Gamma_t\cap\Gamma_\chi=\emptyset$.
  Each material point
  in $\Omega$ is  labeled by its position vector
$\mathbf{X}$ and the corresponding point  in the deformed
configuration is denoted by $\mathbf{x}\,(=\bchi(\mathbf{X}))$. The body
force $\bff$ (per unit mass) is taken to be conservative with
potential $\phi(\mathbf{x})$ so that $\bff=-\grad \phi$, and
$\rho$ is the reference mass density. On the part $\Gamma_t$ of
the boundary the surface traction $\bt$ is prescribed to be of
dead-load type, while on $\Gamma_\chi$ the deformation $\bchi$ is
given. The notation $\calX_a$ identifies a \emph{kinematically
admissible space} of deformations $\bchi$, defined by
\begin{equation}
\calX_a = \{ \bchi \in {\cal W}^{1,p} (\Omega; \real^3)\; \big| \;\ \nabla \bchi
\in \calF_a, \;\; \bchi  =\bchi_0 \;\; \mbox{ on }
\Gamma_\chi \},
\end{equation}
where ${\cal W}^{1,p}$ is the Sobolev space, i.e. a  function space in which both
$\bchi$ and its weak derivative  $\nabla \bchi$ have a finite $L^p(\Oo)$ norm.
${\cal F}_a = \{ \bF \in \calL^p(\Oo; \real^{3 \times 3} )| \; \det \bF > 0 \} $ denotes the admissible deformation gradient space  with $p > 1$. Clearly,  solutions $ \bchi \in \calX_a$ of the problem $(\calP)$ are not necessarily to be smooth.

The criticality condition $\delta \Pi(\bchi) = 0$ leads to a
mixed boundary-value problem $(BV\!P)$, namely
\begin{equation}
(BVP): \;\; \left\{
\begin{array}{l}
\nabla \cdot [\nabla_\bF W(\nabla \bchi) ]  + \rho\bff = \mathbf{0} \quad \mbox{in }
\Omega,\\[0.2cm]
\bn  \cdot    [\nabla_\bF W(\nabla \bchi) ]  = \bt  \quad\mbox{on } \Gamma_t,
\end{array} \right. \label{eq-bvp1}
\end{equation}
where $\nabla_\bF W(\nabla \bchi) = \partial W(\bF)/\partial \bF$
(in components $ \partial W/\partial F_{i\alpha}$), $\bn$ is the unit
outward normal to $\Gamma_t$ and, in component form, we adopt the
conventions $\nabla \cdot \btau= \{\partial \tau_{i\alpha}/\partial
X_\alpha \}$ and $\btau\cdot \bn= \{\tau_{i\alpha}n_\alpha \}$. Note that
$\nabla \cdot \btau$ is defined in the weak sense where
$\nabla\bchi$ is discontinuous. In general, it is rarely possible
to solve this nonlinear boundary-value problem by use of direct
methods. Indeed, the strain energy $W(\bF)$ is a nonconvex
function of $\bF$, the problems $(\calP)$ and $(BVP)$ are not
equivalent, and $(BVP)$ may possess multiple solutions.
Identification of the global minimizer of the variational problem
$(\calP)$ is a fundamentally difficult task in nonconvex analysis.
From the point of view of numerical analysis, any numerical
discretization of the problem $(\calP)$ leads to a nonconvex
minimization problem, and it is well known in global optimization
theory that most nonconvex minimization problems are NP-hard
\cite{gao-jogo00,gao-amma03,gao-optm03}.

Duality principles play  fundamental roles in  sciences and engineering,
especially in continuum mechanics and variational analysis. For linear elasticity,
since  the stored strain energy $W$ is a convex function of the (infinitesimal) strain
 tensor, it is well-known that
each potential   variational (primal) problem is linked a unique  equivalent (dual) complementary
variational problem via the conventional Legendre transformation.
 This one-to-one duality relation is also known as the complementary variational principle,
which has been well-studied with extensive applications in both
mathematical physics and engineering mechanics
(see Arthurs, Nobel-Sewell, Oden-Reddy, Tabarrok-Rimrott, etc).

In finite deformation theory, if the stored energy density $W(\bF)$ is a strictly convex
function of the deformation gradient tensor  $\bF$ over the field $\Oo$,
then the first Piola-Kirchhoff stress tensor
 can be uniquely determined by $\btau = \nabla W(\bF)$ and the complementary energy
  density $W^*$ can be obtained explicitly   by the
 Legendre transformation:
\begin{equation}
W^*(\btau) = \left\{ \bF\! :\! \btau - W(\bF)\, \big | \;\; \btau
= \nabla  W(\bF)  \right\},
\end{equation}
where $\bF  : \btau$ is defined as $\tr(\bF\cdot\btau^{\rm T})$
and $^{\rm T}$ signifies the transpose. In this case,
 the
complementary variational problem can be defined  as
\begin{equation}
\min_{\btau \in \calT_a}  \left\{ \Pi^c(\btau) = \int_\Omega
W^*(\btau) {\rm d} \Omega
- \int_{\Gamma_\chi} \bchi_0 \cdot \btau \cdot \bn {\rm d} \Gamma
\right\} , \label{comprom}
\end{equation}
where $\calT_a$ is the \emph{statically admissible space} defined
by
\begin{equation}
\calT_a = \left\{ \btau \in {\cal L}^q(\Omega)\; \big | \ \nabla
\cdot \btau +  \rho \bff = \mathbf{0}\ \mbox{ in } \Omega, \ \btau
\cdot\bn = \bt\ \mbox{ on } \Gamma_t \right\},
\end{equation}
where $q$ is the conjugate number of $p$, i.e. it is given by $1/p
+ 1/q = 1$.
This complementary variational  problem was
first studied by Levinson \cite{levi-65}.
The well-known Levinson principle states that
if $\barbtau$ is a solution of the complementary variational problem (\ref{comprom}),
then the deformation field
$\barbchi$ defined through the inverse
constitutive law $\bF(\barbchi) = \nabla W^*(\barbtau)$ is a solution of the potential variational problem  (\ref{pprobm})
and the complementarity condition
\[
\Pi(\barbchi) + \Pi^c(\barbtau)  = 0
\]
holds.
This principle can be proved easily by using the traditional Lagrangian duality theory
(see Gao, 2000).

The Levinson  principle is simply the counterpart in finite deformation
theory of the complementary variational principle in linear
elasticity. In finite deformation theory, the stored strain energy
$W(\bF)$ is  in general  nonconvex such that  the stress-deformation
relation $\btau = \nabla W(\bF) $ is not uniquely
invertible \cite{ogden75,ogden77} and the complementary energy
function $W^*$ cannot   be defined explicitly via the
  Legendre transformation. Although   by the Fenchel
transformation
$$
W^\sharp (\btau) = \max_{\bF} \{ \bF\! :\! \btau - W(\bF ) \},
$$
 the Fenchel-Moreau  type dual problem can be formulated in the form of
\begin{equation}
\max_{\btau \in \calT_a} \left\{ \Pi^\sharp (\tau) =
\int_{\Gamma_\chi} \bchi_0 \cdot \btau \cdot \bn {\rm d} \Gamma
-\int_\Omega W^\sharp(\btau) {\rm d} \Omega \right\}, \label{eq-Pic}
\end{equation}
the nonconvexity of  $W$
 leads only to the so-called \emph{weak duality theorem}
$$
  \min_{\bchi \in \calX_a} \Pi(\bchi) \ge  \max_{\btau
\in \calT_a} \Pi^\sharp(\btau)
$$
due to
the Fenchel-Young inequality$
W(\bF)  \ge \bF\! :\! \btau - W^\sharp(\btau)$.
In nonconvex analysis, the nonzero  $\theta = \min_{\bchi \in \calX_a} \Pi(\bchi) -  \max_{\btau
\in \calT_a} \Pi^\sharp(\btau) > 0 $ is called the
\emph{duality gap}. This duality gap shows that the  well-developed
Fenchel-Moreau  duality theory can be used to solve mainly
convex problems.

In finite deformation theory, the well-known Hellinger-Reissner
principle \cite{hell-14, reiss53} and the Fraeijs de Veubeke
principle \cite{veub72} hold for both convex and nonconvex
problems. However, these principles are not considered as
\emph{pure complementary variational principles} since the
Hellinger-Reissner principle involves both the displacement field
and the second Piola-Kirchhoff stress tensor; while the Fraeijs de
Veubeke principle has both the rotation tensor and the first
Piola-Kirchhoff stress as its variational arguments. The existence
of a pure complementary variational principle in general finite
deformation theory has been discussed by many researchers over
several decades (see, for example, \cite{koiter76,
lee-shie80, lee-shie80b, li-cupta,oden-redd83,ogden75,ogden77}).
Moreover, since
the extremality condition in nonconvex variational analysis and
global optimization is fundamentally difficult to resolve, none of
the classical complementary-dual variational principles in finite
deformation theory can be used for reliable numerical
computations.

Canonical duality theory provides a potentially useful
methodology for solving a large class of nonconvex problems in complex systems.
This theory consists mainly of (1) a \emph{canonical dual
transformation}, which can be used
 to formulate  perfect dual problems in nonconvex systems;
 (2) a {\em complementary-dual variational principle,} which
 allows a unified  analytical solution form in terms of the canonical dual solutions;
  (3) a \emph{triality theory},  which provides sufficient criteria for identifying both
global and local extrema.
The original idea of the canonical dual transformation was
introduced by Gao and Strang \cite{gao-strang89a} in finite deformation systems.
In order to recover the duality gap in nonconvex variational problems,
 they discovered  a so-called
 {\em complementary gap function}, which leads to a
 complementary-dual variational principle in finite deformation mechanics.
 They proved that if this gap function is positive on a dual feasible space,
the generalized Hellinger-Reissner energy is a saddle-functional.
It turns out that this gap function provides a sufficient condition  for global optimal solution
in nonconvex variational problems.
Seven years later, it was realized that the negative gap function can be used to identify local extrema.
 Therefore,  a triality theory was first proposed in post-buckling problems of a large deformation beam model
\cite{gao-amr97}, and  a pure complementary energy principle  was eventually obtained in \cite{gao-mrc99}.
This principle can be used to obtain a general analytical solution for 3-D large deformation elasto-plasticity
 \cite{gao-mecc99}.
It was shown by Gao and Ogden (see \cite{gao-ogden-qjmam08,gao-ogden-zamp})
that   for one-dimensional nonlinear elasticity problems,
 both global and local minimal solutions are usually nonsmooth
 and can't be obtained by any Newton type of numerical methods.
For finite dimensional systems, the
 canonical duality theory has been successfully applied for solving a large class of challenging
  problems in computational mechanics \cite{cai-gao-qin,gao-yu,hugo-gao}
  and global optimization with extensive applications in computational biology \cite{zgy},
  chaotic dynamical systems \cite{li-zhou-gao,ruan-gao-ima},
   discrete  and network optimization \cite{gao-cace09,gao-ruan-jogo08,gao-ruan-pardalos,ruan-gao-jiao-coap08}.

The purpose of this paper is to illustrate the application of the
pure complementary variational principle in combination with
triality theory by solving a  general nonconvex variational
problem governed by St Venant-Kirchhoff material. The paper is organized as follows.  Section
\ref{primal}  presents a brief review on  the canonical duality theory
in nonlinear elasticity.  Some fundamental issues in nonlinear elasticity are addressed, including
the reasons why the Legendre-Hadamard
condition provides   only necessary condition for local minima,
how the Gao-Strang gap function and the triality theory can be used to identify both global and local
extremal solutions.
  In Section
\ref{stvenant} we show that for the St Venant-Kirchhoff materials,
 the pure complementary variational problem
can be solved  principally  to obtain all possible solutions.
 Some concluding remarks are contained in Section \ref{finish}.

\section{Canonical Duality Theory and Complementary Variational Principle}\label{primal}
It is known that the stored-energy function $W:\calF_a \rightarrow \real$ must obey certain
physical laws and requirements in continuum mechanics, such as the  principle of material frame-indifference \cite{Truesdell-Noll},
which lay a mathematical foundation for the
canonical duality theory. Let
\eb
\SO(3) = \{ \bQ \in \real^{3\times 3} | \; \bQ^T = \bQ^{-1} , \;\; \det \bQ = 1 \}
\ee
be the special orthogonal group.

\begin{definition}[Objectivity and Isotropy \cite{gao-dual00}] $\;$\newline \vspace{-.5cm}
\begin{verse}
{\em
(D1) {\em Objective Set and Objective Function}: A subset $\calF_a \subset \real^{3\times 3}$ is
said to be {\em objective} if for every $\bF \in \calF_a$ and every $\bQ \in \SO(3)$,
$\bQ \bF \in \calF_a$.
A scalar-valued function $W:\calF_a \rightarrow \real$ is said to be {\em  objective }
if its domain is objective and
\eb
W(\bQ \bF) = W(\bF) \;\; \forall \bF \in \calF_a, \; \forall \bQ \in \SO(3).
\ee

(D2) {\em Isotropic Set and Isotropic Function}: A subset $\calF_a \subset \real^{3\times 3}$ is
said to be {\em isotropic} if for every $\bF \in \calF_a$ and every $\bQ \in \SO(3)$,
$ \bF\bQ \in \calF_a$.
A scalar-valued function $W:\calF_a \rightarrow \real$ is said to be {\em  isotropic }
if its domain is isotropic and
\eb
W( \bF\bQ) = W(\bF) \;\; \forall \bF \in \calF_a, \; \forall \bQ \in \SO(3).
\ee
}
\end{verse}
\end{definition}

The objectivity   implies that the constitutive law of material is independent with the observer (coordinate free).
While the isotropy means that the material possesses certain symmetry.
Generally speaking, the deformation gradient $\bF$ is a two-point tensor, which is not considered as
a strain measure.
The {\em right Cauchy-Green tensor } $\bC = \bF^T \bF$ is a (Lagrange type) strain measure
which is  objective (rotation free), i.e.,
\[
\bC (\bQ \bF)= (\bQ \bF)^T (\bQ \bF) = \bF^T \bQ^T \bQ \bF = \bC(\bF)  \;\; \forall \bQ \in \SO(3).
\]
Dually,  the {\em left Cauchy-Green tensor} $\bB = \bF^T \bF$ is an isotropic function of $\bF$.

In continuum mechanics, the objectivity is also known as   {\em the principle of  frame-indifference}.
According to P.G. Ciarlet, the stored energy function of a hyper-elastic material is objective if and
only if there exists a function $\UU(\bC)$ such that $W(\bF) = \UU(\bC(\bF))$ (see Theorem 4.2-1 in \cite{ciarlet}).
 This  principle lays a foundation for the canonical duality theory.

 Indeed, the canonical dual transformation was developed from the concept of the objectivity.
The key step of  this   transformation   is the introduction of a  geometrically admissible strain  measure
$\bveps = \bLam (\bchi):\calX_a \rightarrow \calE_a \subset \real^{3\times 3}$  and the  {canonical function}
$U(\bveps) : \calE_a \rightarrow \real $ such that the nonconvex stored energy  $W(\bF)$ can be written in the canonical form
$W(\nabla \bchi ) = U(\bLam(\bchi))$.
According to  \cite{gao-dual00},
 a convex differentiable real-valued function $\UU(\bxi) $
is said to be canonical on its domain $\calE_a$ if the duality relation
$\bxi^*  = \nabla \UU(\bxi) : \calE_a \rightarrow \calE_a^*  $ is invertible
 such that  the conjugate
function $U^*(\bxi^*)$ of $U(\bveps)$ can be defined uniquely by the Legendre
transformation
\eb
\UU^*(\bxi^*) = \{ \bxi : \bxi^* - \UU(\bxi) | \; \bxi^* = \nabla \UU(\bxi) \;\; \forall \bxi \in \calE_a \}.
\ee
By the theory of convex analysis, it is easy to prove  that the following
canonical duality relations hold on $\calE_a \times \calE^*_a$
\eb
\bxi^* = \nabla \UU(\bxi) \; \Leftrightarrow \;\; \bxi = \nabla \UU^*(\bxi^*) \; \Leftrightarrow \; \UU(\bxi) + \UU^*(\bxi^*)
= \bxi : \bxi^*
\ee
and the  pair $(\bxi, \bxi^*)$  is called the {\em canonical dual pair } on $\calE_a \times \calE_a^*$.

Thus, on replacing  $W(\nabla \bchi)$ in the total
potential energy $\Pi(\bchi)$ by its  canonical form
$W(\nabla \bchi) = U(\bLam(\bchi))$,   and we  take the body force to be a constant, so that
$\phi(\bchi)=-\mathbf{f}\cdot\bchi$,
the minimal potential energy variational problem (1) can be written in the following canonical form
\eb
(\calP): \;\; \min_{\bchi \in \calX_a}  \left \{
\Pi(\bchi) = \int_\Oo [U(\bLam(\bchi)) -   \rho  \bchi \cdot \mathbf{f}]  \dO - \int_\Gt \bchi \cdot \bt \dG \right\}.
\ee

Furthermore, in terms of $\bvsig = \bxi^*$ and by  the Fenchel-Young equality
\[
U(\bLam(\bchi)) = \bLam(\bchi)\! :\! \bvsig - U^*(\bvsig),
\]
 the so-called
\emph{total complementary energy functional} \cite{gao-strang89a} $\Xi: \calX_a \times \calE^*_a \rightarrow \real$
can be written, in the present
context, as
\begin{equation}
\Xi(\bchi, \bvsig) = \int_{\Omega} \left[\bLam(\bchi)\! :\! \bvsig
- U^*(\bvsig) - \rho  \bchi \cdot \mathbf{f}  \right] {\rm d} \Omega -
\int_{\Gamma_t} \bchi \cdot \bt {\rm d} \Gamma.
\end{equation}
For a given statically admissible field $\btau \in \calT_a$, this total complementary functional can be written
in the following form
\eb
\Xi_\btau (\bchi, \bvsig) = \int_{\Gamma_\chi} \bchi_0 \cdot \btau \cdot \bn {\rm d} \Gamma
+ \int_{\Omega} \left[\bLam(\bchi)\! :\! \bvsig
- U^*(\bvsig) -  (\nabla \bchi): \btau \right] \dO .
\ee
  For a given $\bvsig \in \calE^*_a$, the
\emph{canonical dual functional} $\Pi^d(\bvsig)$ is then defined
by
\begin{equation}
\Pi^d(\bvsig) = \left\{ \Xi(\bchi, \bvsig) \;\big | \;
\delta_{\bchi} \Xi(\bchi, \bvsig) =0\right\}   =
 F^{\bLam}(\bvsig) - \int_\Omega U^*(\bvsig) {\rm d} \Omega,
\end{equation}
where $F^{\bLam}(\bvsig)$ is defined by the so-called $\bLam$-conjugate
transformation \cite{gao-dual00, gao-optm03}
\begin{equation}
F^{\bLam}(\bvsig) =\sta \left\{ \int_{\Omega} [\bLam(\bchi)\! :\!
\bvsig - \rho  \bchi \cdot \mathbf{f} ] {\rm d} \Omega - \int_{\Gamma_t} \bchi
\cdot \bt {\rm d} \Gamma \; \big| \;\;  \bchi \in \calX_a
\right\},
\end{equation}
with sta indicating the stationary value at fixed $\bvsig \in \calE^*_a$.
 In terms of $\btau \in \calT_a$, we have the following form
\begin{equation}
F^{\bLam}_\btau (\bvsig) = \int_{\Gamma_\chi} \bchi_0 \cdot \btau \cdot \bn {\rm d} \Gamma
+ \sta \left\{ \int_{\Omega} [\bLam(\bchi)\! :\!
\bvsig - (\nabla \bchi ): \btau   ] {\rm d} \Omega
 \; \big| \;\;  \bchi \in \calX_a
\right\}.
\end{equation}
In finite deformation theory,
\eb
 \Pi^d_\btau(\bvsig) = F^{\bLam}_\btau (\bvsig) - \int_\Oo  \UU^*(\bvsig) \dO
 \ee
 is also called the
\emph{pure complementary energy functional},  which was first proposed in  \cite{gao-mrc99}.
\begin{thm}[Complementary-Dual Variational Principle \cite{gao-mecc99}]
For a given statically admissible field $\btau \in \calT_a$, the following statements are equivalent:
\begin{enumerate}
  \item $(\barbchi, \barbvsig)$
is a critical point of $\Xi_\btau(\bchi, \bvsig)$;
  \item  $\barbchi$ is a critical point of   $\Pi(\bchi)$;
  \item $\barbvsig$ is a critical point of $\Pi^d_\btau(\bvsig)$.
\end{enumerate}

Moreover, we have
\eb
\Pi(\barbchi) = \Xi (\barbchi, \barbvsig) = \Xi_\btau(\barbchi, \barbvsig)= \Pi^d_\btau(\barbvsig).
\ee
\end{thm}

This theorem shows that to find a critical solution to the nonconvex total potential $\Pi(\bchi)$
is equivalent to find
a critical point of its  canonical dual function $\Pi^d_\btau (\bvsig)$.
For a given $ \btau \in \calT_a$,
different choice of  the geometrical measure $\bLam(\bchi)$   will leads to   different,  but equivalent,
 $\Pi^d_\btau(\bvsig)$  on a subset $\calS_a \subset \calE^*_a$.

 In finite deformation theory, the canonical duality relation is also known as  the Hill  {\em work conjugate}
 and  the canonical function $\UU(\bxi)$ is called strain energy-density.
 According to Hill,  for a given hyper-elastic material,
there  exist a class of strain measures $\bxi$  and the associated  canonical functions $\UU(\bxi)$ such that
 the associated stress can by defined uniquely by the canonical duality relation $\bxi^* = \nabla \UU(\bxi)$.
 There are many canonical strain measures in finite elasticity and many of these
strain measures   belong to the  well-known Hill-Seth strain family
\[
\bE^{(\eta)} = \frac{1}{2 \eta} [ \bC^{\eta} - \bI ],
\]
where $\bI $ is an identity tensor in $\real^{3\times 3}$ and $\eta $ is a real number.

Canonical duality theory and pure complementary energy principle for general strain measures have
 been studied in \cite{gao-dual00}.
In this paper, we consider only the  Green-St Venant  strain tensor $\bE^{(1)}$,
simply denoted as $\bE $.
In this case, the geometrical operator
\eb
\bE = \Lam(\bchi) =   \half [ (\nabla \bchi)^T (\nabla \bchi)  - \bI ] : \calX_a \rightarrow \calE_a
 \ee
 is a quadratic operator and its domain can be defined by
\eb
\calE_a = \{ \bE \in  \calL^{p/2}(\Oo; \real^{3 \times 3}) | \; \bE = \bE^T,
\;    (2 \bE + \bI)   \succ   0 \}.
\ee

We assume that the associated strain energy density $\UU(\bE):\calE_a \rightarrow \real$ is
convex such that the conjugate stress $\bvsig$ of $\bE$, denoted by $\bT$, can be defined uniquely by the
constitutive law
\eb
\bT = \nabla \UU(\bE) :\calE_a \rightarrow \calE^*_a .
\ee
  This associated stress $\bT $ is the well-known second Piola-Kirchhoff stress, which is
well-defined on $\calE^*_a = \{ \bT \in \calL^{p/(p-2)}(\Oo; \real^{3\times 3} )| \;\; \bT = \bT^T \}$.
    In this case, the   pure complementary energy $\Pi^d_\btau $ has the form of
\begin{equation}
\Pi^d_\btau (\bT) = \int_{\Gamma_\chi} \bchi_0 \cdot \btau \cdot \bn
{\rm d} \Gamma -\int_\Omega \left[ \frac{1}{2}\tr
( \btau \cdot \bT^{-1}
\cdot \btau + \bT ) +   U^*(\bT) \right]{\rm d} \Omega ,
\end{equation}
which is well-defined on
the canonical dual space
\eb
\calS_a = \{ \bT \in \calE^*_a | \;\; \tr( \btau \cdot \bT^{-1}  \cdot \btau)  \in \calL^1(\Oo; \real)\; 
\; \forall \btau \in \calT_a \}.
\ee
Therefore, the canonical dual problem  is to find the critical point $\barbT \in \calS_a$ such that
\eb
(\calP^d): \;\; \Pi^d_\btau(\barbT) = \sta \{ \Pi^d_\btau(\bT) | \; \bT \in \calS_a \} .
\ee

\begin{thm}[Analytical Solution Form \cite{gao-dual00}]\label{thm-ana}
For a given $\btau \in \calT_a$, if  $ \barbT $ is a critical point of
  $\Pi^d_\btau(\bT)$,
 then along any path from
$\mathbf{X}_0 \in \Gamma_\chi$ to $\mathbf{X} \in \Omega$, the
deformation defined by
\begin{equation}
 \bar{\bchi} = \int_{\mathbf{X}_0}^{\mathbf{X}}   \btau \cdot \barbT^{-1} \cdot
{\rm d} \mathbf{X} + \bchi_0(\mathbf{X}_0) \label{eq-anasolu}
\end{equation}
is a critical solution to $(\calP)$.
Moreover, if
  \eb
  \nabla \times (\btau \cdot
\barbT^{-1}) = \mathbf{0}, \label{eq-compat}
\ee
then $\bar{\bchi} $
is a closed form solution to the boundary value problem (BVP)
(\ref{eq-bvp1}).
\end{thm}

The proof of this theorem can be found in  \cite{gao-ima98,gao-mrc99,gao-mecc99}.
In fact,  the criticality condition $\delta
\Pi^d_\btau(\bT)= 0$ leads to  the following  dual tensor equation:
\begin{equation}
\bT \cdot \left[ \bI + 2  (\nabla  U^*(\bT)) \right] \cdot \bT =
 {\btau}^{\rm T} \cdot\btau , \label{cdtevk}
\end{equation}
which is equivalent to
\[
\nabla \UU^* (\barbT) = \half \left( (\btau \cdot \barbT^{-1})^T \btau \cdot \barbT^{-1} - \bI  \right) .
\]
This is actually the constitutive law $\bE = \bLam(\barbchi) = \half [\bF^T \bF - \bI] = \nabla \UU^*(\barbT)$
subjected to $\bF = \btau \cdot \barbT^{-1}$.
Therefore, if   the compatibility condition
$\nabla\times\bF=\mathbf{0}$, in index notation
\[
\frac{\partial F_{i\alpha}}{\partial X_\beta } = \frac{\partial F_{i\beta}}{ \partial
X_\alpha} ,
\]
 holds, then  $\bF$ is the deformation  gradient and $\barbchi$ is a solution to $(BVP)$.
%

\begin{rem}[PDE $\Leftrightarrow$ Algebraic Equation]
{ Theorem \ref{thm-ana} shows that by the pure complementary energy principle, the
nonlinear partial differential equation $(BVP)$ is  equivalently converted to a canonical dual tensor equation (\ref{cdtevk}),
which can be solved to obtain the stress field $\barbT$ for certain materials.
From the equation (\ref{cdtevk}) we know that  $\bT = {\bf 0} $ if $\btau = {\bf 0}$. Therefore, although
 $\bT^{-1}$ appears in $\Pi^d_{\btau}(\bT)$,  this pure complementary energy   is well-defined on  $\calS_a$.
The equation (\ref{eq-anasolu})
presents an analytical solution form to the boundary value problem in terms of
the canonical dual stress field $\barbT$ and the statically admissible $\btau \in \calT_a$.
Of  course, this is purely formal and in general it is not easy to obtain the solution for general practices
unless the deformation compatibility condition (\ref{eq-compat}) holds.

It has been assumed here that the relation between $\bT$ and
$\bE$ is invertible.  This certainly holds in a neighborhood of
the (stress-free) reference configuration since the canonical strain energy $\UU(\bE)$
is convex  in such a neighborhood.  It is a reasonable
assumption to extend this to a sufficiently large domain that
includes deformations of practical interest.  Finite element
implementations of nonlinear elasticity are usually based on the
variables $\bT$ and $\bE$ and the associated tangent tensor
$\partial\bT/\partial\bE = \nabla^2 \UU(\bE)$, which is assumed to be positive
definite.  It is always possible to select forms of the
strain-energy function $W$ such that this is the case, although
the possibility of its failure for particular materials is not in
general ruled out.}
\end{rem}

\medskip
In terms of the deformation $\bchi \in \calX_a$ and the second Piola-Kirchhoff stress $\bT \in \calE^*_a$,
the total complementary functional  $\Xi(\bchi,\bT)$  can be written as
\eb
\Xi_\btau(\bchi, \bT) =  \int_{\Omega} \left[\bE(\bchi)\! :\! \bT
- U^*(\bT) - (\nabla  \bchi) : \btau  \right] {\rm d} \Omega
+ \int_{\Gamma_\chi} \bchi_0 \cdot \btau \cdot \bn {\rm d} \Gamma
\end{equation}
which is actually the  well-known Hellinger-Reissner energy if the first Piola-Kirchhoff stress is replaced by
external force field.
From the  nonlinear  canonical dual tensor equation (\ref{cdtevk}) we know that
for a given   $\btau \in \calT_a$, the
pure complementary energy $\Pi^d_\btau(\bT)$ may have multiple critical points.
 In order to identify the global extremum,
We need to introduce the following subspaces:
\eb
\calS^+_a = \{ \bT \in \calS_a | \;\; \bT \succ 0 \}, \;\;
\calS^-_a = \{ \bT \in \calS_a | \;\; \bT \prec 0 \}.
\ee
\begin{thm} \label{thm-tri}
Suppose for a given $\btau \in \calT_a$, the pair  $(\barbchi, \barbT)$ is an isolated   critical point of $\Xi_\btau(\bchi, \bT)$.
If $\barbT \in \calS^+_a$, then $\barbchi$ is a global minimizer of $\Pi(\bchi)$ on $ \calX_a$ if and only if
$\barbT $ is a global maximizer of $\Pi^d_\btau(\bT)$ on $\calS^+_a$, i.e.,
\eb \label{sadminmax}
\Pi(\barbchi) = \min_{\bchi \in \calX_a} \Pi(\bchi)
\;\;\Leftrightarrow \;\; \max_{\bT \in \calS^+_a} \Pi^d_\btau(\bT) = \Pi^d_\btau(\barbT).
\ee

If $\barbT \in \calS^-_a$, then $\barbchi$ is a local maximizer of $\Pi(\bchi)$
if and only if $\barbT$ is a local maximizer of $\Pi^d_\btau(\bT)$, i.e., on a neighborhood $\calX_o \times \calS_o
\subset \calX_a \times \calS^-_a$,
\eb
\Pi(\barbchi) = \max_{\bchi \in \calX_o} \Pi(\bchi)
\;\;\Leftrightarrow \;\;
 \max_{\bT \in \calS_o} \Pi^d_\btau(\bT) = \Pi^d_\btau(\barbT). \label{eq-dobmax}
\ee

If $\barbT \in \calS^-_a$ and $\nabla^2_{\bF} \WW(\nabla \barbchi) \succ  0 $, then $ \barbchi$  is a local minimizer of $\Pi(\bchi)$.
\end{thm}

\begin{rem}[The Complementary Gap Function and Triality Theory] $\;$ \newline
Theorem \ref{thm-tri} shows that the extremality of the primal solution $\bchi$ depends on its canonical dual solution $\bS$.
This result was first discovered by Gao and Strang in 1989 \cite{gao-strang89a}, i.e. they proved that
$\barbchi (\barbS)$ is a global minimizer of
$\Pi(\bchi)$ if the complementary gap function satisfies
\eb
G_{ap}(\bchi,\barbS) = \int_\Oo \half  [(\nabla \bchi)^T (\nabla\bchi) + \bI]:\barbS \dO \ge 0 \;\;
\forall \bchi \in \calX_a \label{eq-gapp}
\ee
Since $G_{ap}(\bchi,\barbS)$  is quadratic in $\bchi$, this gap function is positive
for any given $\bchi \in \calX_a$ if $\barbS \succeq 0 $.
Replacing $\bF = \nabla \bchi$ by $\bF = \btau \cdot \bS^{-1}$, this gap function can be written as the so-called
pure gap function
\eb
G_{ap}(\bchi(\bS), \bS) = \int_\Oo \half \tr (\btau \cdot \bS^{-1} \cdot \btau + \bS) \dO ,
\ee
which is a main term in the pure complementary energy $\Pi^d_\btau(\bS) $ in addition to $\UU^*(\bS)$.
 Comparing $\Pi^d_\btau(\bS) $ with   $\Pi^\sharp(\btau)$ given by (\ref{eq-Pic}),
we can understand that this gap function not only recovers the duality gap in the Fenchel-Moreau duality theory,
but also provides a global extremality condition for nonconvex variational problem $(\calP)$.

 To see this in  detail, let us consider the    canonical transformation  $W(\bF) = \UU(\bE(\bF))$. By  chain rule we have
 \eb
 \frac{\partial^2 W(\bF)}{\partial F^i_\alp \partial F^j_\beta} = \delta^{ij}  S_{\alp\beta} +
 \sum_{\theta, \nu = 1}^3  F^i_\theta  H_{\theta \alp\beta \nu} F^j_\nu, \label{eq-hessian}
 \ee
 where ${\bf H} = \{ H_{\theta \alp\beta \nu}\} = \nabla^2 \UU(\bE)$. By the convexity of  the canonical function $\UU(\bE)$,
we have  ${\bf H} \succ 0 $.  Therefore, if
  $\bS = \{ S_{\alp\beta} \}\in  \calS_a^+ $, the Hessian $\nabla^2 \WW(\bF) \succ 0$ and,  by Gao and Strang  \cite{gao-strang89a},
  the associated deformation field $\bchi$ is a
  global minimizer of $\Pi(\bchi)$.
 The statement (\ref{sadminmax}) shows that the nonconvex minimization problem $(\calP)$ is equivalent to
a concave maximization dual problem over a convex space $\calS^+_a$, i.e.,
\eb
\max \{ \Pi^d_\btau(\bT) | \;\; \bT \in \calS_a^+ \},
\ee
which is much easier than the nonconvex primal problem $(\calP)$.
The global optimality condition $\bS \in \calS_a^+$ is a strong case of
 Gao and Strang's  positive gap function  (\ref{eq-gapp}).

 Subsequently, in a study of
post-buckling analysis for a nonlinear beam theory, it was found
that if the dual solution ${\bar{\bT}} $ is negative definite in
the domain $\Omega$,   the solution $\bar{\bchi}$ could be either a local
minimizer or a local maximizer of the total potential energy. To see this,
we substitutive  $\bF = \btau \cdot  \bT^{-1}$ into (\ref{eq-hessian}) to obtain
  \eb
     \frac{\partial^2 \WW(\bF)}{\partial F^i_\alp \partial F^j_\beta} =  \delta^{ij}  S_{\alp\beta} +
 \sum_{\theta, \nu, \delta, \lambda = 1}^3  \tau^i_\theta S^{-1}_{\theta \delta}  H_{\delta \alp\beta \nu} S^{-1}_{\nu\lambda}\tau^j_{\lambda}
 \ee
which shows that  even if $\bT \prec 0$,
 the Hessian matrix
$\nabla^2 \WW(\bF)$  could be either positive or negative definite, depending on the  eigenvalues of $\bT  \in \calS^-_a$.
Thus,  in addition to
the double-max duality (\ref{eq-dobmax}), we have the so-called double-min duality
\eb
\Pi(\barbchi) = \min_{\bchi \in \calX_o} \Pi(\bchi)
\;\;\Leftrightarrow \;\;
 \min_{\bT \in \calS_o} \Pi^d_\btau(\bT) = \Pi^d_\btau(\barbT),
\ee
which holds under certain condition (see \cite{gao-amma03}).
For
this reason, a so-called triality theory was proposed  first in post-buckling analysis of a large deformed beam model \cite{gao-amr97},
and then in general nonconvex mechanics \cite{gao-mecc99,gao-dual00}. This triality theory reveals an important fact in nonconvex analysis,
i.e. for a given statically admissible field $\btau \in \calT_a$, if the canonical dual equation
(\ref{cdtevk}) has multiple solutions $\{\bS_k\}$ in a subset $\Omega_o \subset \Omega$, then the
boundary value problem $(BVP)$ could have an infinite number of solutions $\{ \bchi_k (\bX) \}$ in $\Omega$.
The well-known Legendre-Hadamard  (L-H) condition   is only a  necessary condition for a local minimal solution,
while the triality theory can identify not only
 the  global minimizers, but also both local minimizers and local maximizers.
  It is known that an elliptic  equation is corresponding to a convex variational problem.
If  the boundary-value problem (\ref{eq-bvp1}) has multiple solutions $\{ \bchi_k (\bX) \}$
at one material point  $\bX \in \Oo$, the total potential $\Pi(\bchi)$ is not convex and the operator
 $A(\bchi) = \nabla \cdot [\nabla_\bF W(\nabla \bchi) ] $ may not be elliptic at   $\bX \in \Oo$
 even if the L-H  condition holds  at certain  $\bchi_k (\bX) $.
\end{rem}

  The pure complementary energy
principle and triality theory play a fundamental role not only in nonconvex
 analysis, but also in computational science and global optimization (see
\cite{gao-amma03,gao-mms04,gao-cace09,gao-review14}).

\section{Application to St Venant-Kirchhoff Material}\label{stvenant}
For St. Venant-Kirchhoff material, the canonical energy function $\UU(\bE)$ has the most simple form:
\eb
U(\mathbf{E})=\mu\tr(\mathbf{E}^2)+\frac{1}{2}\lambda(\tr\mathbf{E})^2.
\ee
The second Piola-Kirchhoff stress depends linearly on the Green-St Venant strain via the Hooke's law:
\eb
\bS = \nabla U(\bE) =
2\mu \bE +\lambda(\tr\bE )\mathbf{I} = \bH  : \bE ,
\ee
where $\bH$ is the Hooke tensor for St Venant-Kirchhoff material.
The complementary energy is
\eb
U^*(\bS)=\frac{1}{4\mu}\tr(\bS^2)-\frac{\lambda}{4\mu(3\lambda+2\mu)}(\tr\bS)^2,
\ee
and hence
\eb
\bE = \nabla \UU^*(\bT)   =
\frac{1}{2\mu}\bT -\frac{\lambda}{2\mu(3\lambda+2\mu)}(\tr\bT)\mathbf{I}\equiv \bH^{-1}:\bS .
\ee
By the canonical dual tensor equation (\ref{cdtevk}), we have
\eb
 {\bS}^2+2\bT( \bH^{-1}: \bT)\bT = \bT^2+\frac{1}{\mu}\bT^3
-\frac{\lambda}{\mu(3\lambda+2\mu)}(\tr\bT)\bT^2
= \btau^T \btau.
\ee
The diagonalization of this tensor equation leads to the following {coupled} quebec nonlinear
algebraic systems:
\eb
S_i^2+\frac{1}{\mu} S_i^3-\frac{\lambda}{\mu(3\lambda+2\mu)}(S_1+ S_2+ S_3) S_i^2 = \tau_i^2
\;\; \quad i=1,2,3 . \label{mainsystem}
\ee
 For convenience, we make the following substitutions in (\ref{mainsystem}):
\[
S_i=\mu \vsig_i, \;\; \tau_i^2=\mu^2 \sig_i, \; \; i=1,2,3,
\]
 and $k = \frac{\lam}{3 \lam + 2 \mu} < 1/3$ (due to $\mu > 0$).
 So, the system (\ref{mainsystem}) can be written as follows
\eb \label{mainsystem1}
\vsig_i^3+ \vsig_i^2- k (\vsig_1+ \vsig_2+ \vsig_3) \vsig_i^2 =  \sig_i ,  \;\; i=1,2,3.
\ee

\subsection{Auxiliary Equation}

In this section we will study solutions of the following equation:
\begin{equation}\label{meq}
    G(\vsig,\qq,\sig)=\vsig^3 + (1-k \qq )\vsig^2 - \sig=0 ,
\end{equation}
 where $\sig > 0$, $0< k <\frac{1}{3}$, and $\qq$ is an arbitrary real number. Also, since $\sig>0$, we can assume that $\vsig\neq 0$.

Since the parameter $q$ in this section is assumed to be independent on $\vsig$, the following results are similar to
one-dimensional nonlinear elasticity problems studied by Gao \cite{ gao-dual00,gao-na00},  Gao and Ogden
\cite{gao-ogden-qjmam08}.

\begin{lem}\label{l1}If $\vsig_1,\vsig_2,\vsig_3$ are solutions of the equations  $G(\vsig,\qq,\sig_1)=0$, $G(\vsig,\qq,\sig_2)=0$, $G(\vsig,\qq,\sig_3)=0$ correspondingly, and $\vsig_1+\vsig_2+\vsig_3=\qq$, then $\vsig_1,\vsig_2,\vsig_3$ satisfy (\ref{mainsystem1}).
\end{lem}
\noindent {\bfseries Proof}.   Obvious.  \hfill $\Box$

\begin{lem}\label{l2}Equation (\ref{meq}) has exactly one positive solution. It has negative solutions iff $$\qq\leq \frac{1}{k}(1-3\sqrt[3]{\frac{\sig}{4}})$$ There is only one negative solution if and only if $\qq= \frac{1}{k}(1-3\sqrt[3]{\frac{\sig}{4}})$.
\end{lem}
\noindent {\bfseries Proof}.  To check that there is exactly one positive root one can apply the Descartes' rule of signs. To prove the rest, let's fix $\qq,\sig$ and notice that $G(\vsig,\qq,\sig)=0$ has negative solutions iff
it has at least two different solutions. This will happen iff the values of the function at local minimum and maximum have different signs.
The extremums of G are at $\vsig_0=-\frac{2}{3}(1-k\qq)$ and 0. Since the value of G at 0 is $-\sig < 0$, we find when $G(\vsig_0,\qq,\sig)\geq 0$. Solving this inequality we get $\qq\leq \frac{1}{k}(1-3\sqrt[3]{\frac{\sig}{4}})$.

\begin{cor}\label{cor2l2} The equation $G(\vsig,0,\sig)=0$ has negative solution(s) iff

\[
\sig\leq \frac{4}{27} .
\]
\end{cor}
\noindent {\bfseries Proof}.  Apply Lemma \ref{l2} to $\qq=0$.  \hfill $\Box$

\begin{lem}\label{l3} Let's fix $\sig>0$ and assume that $\vsig_0,\qq_0$ satisfy (\ref{meq}), and $\vsig_0\neq 0$, $\vsig_0\neq -\sqrt[3]{2\sig}$. Then there exists a unique continuously differentiable function $\vsig(\qq)$, such that $\vsig(\qq_0)=\vsig_0$, $\vsig(\qq)$ and $\qq$ both satisfy (\ref{meq}) and

$$\frac{d\vsig}{d\qq}=\frac{k\vsig^3}{\vsig^3+2\sig} .$$

\noindent Moreover, there are three possibilities (``branches") for $\vsig(\qq)$:

\begin{verse}
  (a)  If $\vsig_0 \in (-\infty,-\sqrt[3]{2\sig})$, then the range of $\vsig(\qq)$ is $(-\infty,-\sqrt[3]{2\sig})$, the domain is $(-\infty, \frac{1}{k}(1-3\sqrt[3]{\frac{\sig}{4}}))$, and $\vsig(\qq)$ is monotonically increasing.

  (b)  If $\vsig_0 \in (-\sqrt[3]{2\sig},0)$, then the range of $\vsig(\qq)$ is $(-\sqrt[3]{2\sig},0)$, the domain is $(-\infty,  \frac{1}{k}(1-3\sqrt[3]{\frac{\sig}{4}})))$ , and $\vsig(\qq)$ is monotonically decreasing.

  (c)  If $\vsig_0 \in (0,+\infty)$, then the range of $\vsig(\qq)$ is $(0,+\infty)$, the domain is $(-\infty, +\infty)$, and $\vsig(\qq)$ is monotonically increasing.
\end{verse}
\end{lem}
\noindent {\bfseries Proof}.  Let's fix $\sig$ and find $\qq$ from (\ref{meq})
$$\qq(\vsig)=\frac{\vsig^3+\vsig^2-\sig}{k\vsig^2}.$$
\noindent Since, $\frac{d\qq}{d\vsig}=\frac{\vsig^3+2\sig}{k\vsig^3}$ and $\sig>0$ it is obvious that $\qq(\vsig)$ is monotonically increasing in the intervals $\vsig\in (-\infty,-\sqrt[3]{2t})$ and $\vsig\in (0,+\infty)$ and is monotonically decreasing in the interval $\vsig\in (-\sqrt[3]{2t},0)$. The corresponding intervals for
$\qq$ are $(-\infty, \frac{1}{k}(1-3\sqrt[3]{\frac{\sig}{4}}))$, $(-\infty, +\infty)$, and $(-\infty, \frac{1}{k}(1-3\sqrt[3]{\frac{\sig}{4}}))$.
  Also, one can easily check that $$\frac{d\vsig}{d\qq}=\frac{k\vsig^3}{\vsig^3+2\sig}.$$
  Thus, the lemma is proved.  \hfill $\Box$

\begin{defi} The three branches of $\vsig(\qq,\sig)$ ($\sig$ is fixed) described in Lemma \ref{l3} will be denoted as follows:
\begin{description}
  \item[(a)]  $\vsig^1(\qq,\sig)$ is a positive branch with the domain $(-\infty, +\infty)$ and range $(0,+\infty)$;
  \item[(b)] $\vsig^3(\qq,\sig)<\vsig^2(\qq,\sig)$ are two negative branches with the domain $(-\infty,  \frac{1}{k}(1-3\sqrt[3]{\frac{\sig}{4}}))$ and ranges $(-\infty,-\sqrt[3]{2\sig})$ and $(-\sqrt[3]{2\sig},0)$ correspondingly.(Note that Corollary 3.1 implies that $\frac{1}{k}(1-3\sqrt[3]{\frac{\sig}{4}})\leq 0$)
\end{description}
\end{defi}

\begin{defi} Let's introduce the following notations:
\begin{description}
  \item[(a)] $\bar{\vsig}^i(\qq,\sig)=\vsig^i(\qq,\sig)-\frac{\qq}{3}$, $i=1,2,3$;
  \item[(b)] $F^{i,j,k}(\qq,\sig_1,\sig_2,\sig_3)= \bar{\vsig}^i(\qq,\sig_1)+\bar{\vsig}^j(\qq,\sig_2)+\bar{\vsig}^k(\qq,\sig_3)$, $i,j,k=1,2,3$.
\end{description}
\end{defi}

\begin{lem}\label{l4}The following statements are true:
\begin{description}
  \item[(a)] For $i=1,2,3$ $$\bar{\vsig}^i(\qq,\sig)=-\frac{(1-3k)\vsig^i(\qq,\sig)^3+\vsig^i(\qq,\sig)^2-\sig}{3k\vsig^i(\qq,\sig)^2}$$
  and
$$\frac{d\bar{\vsig}^i}{d\qq}=-\frac{(1-3k)\vsig^i(\qq,\sig)^3+2\sig}{3(\vsig^i(\qq,\sig)^3+2\sig)} .$$
  \item[(b)] $\vsig^1(0,\sig)=\bar{\vsig}^1(0,\sig)>0$, $\vsig^2(0,\sig)=\bar{\vsig}^2(0,\sig)<0$, and $\vsig^3(0,\sig)=\bar{\vsig}^3(0,\sig)<0$.
  \item[(c)] For a fixed $\sig$, $\bar{\vsig}^1(\qq,\sig)$ is monotonically decreasing in $\qq$ and \\$\lim_{\qq\rightarrow +\infty}\bar{\vsig}^1(\qq,\sig) = -\infty$.
  \item[(d)] For a fixed $\sig$,
  \[
  \lim_{\qq\rightarrow -\infty}\bar{\vsig}^2(\qq,\sig) = +\infty
  \;\;\;
 \mbox{ and } \;\;
      \lim_{\qq\rightarrow -\infty}\bar{\vsig}^3(\qq,t) = +\infty .
      \]
  \item[(e)] For fixed $\sig_1,\sig_2,\sig_3$, each of $F^{i,j,k}(\qq,\sig_1,\sig_2,\sig_3)$, $i,j,k=1,2,3$, is continuous. Moreover, $F^{1,1,1}(\qq,\sig_1,\sig_2,\sig_3)$ is monotonically decreasing in $\qq$.
\end{description}
\end{lem}
\noindent {\bfseries Proof}. To check (a), first substitute $\qq(\vsig)=\frac{\vsig^3+\vsig^2-\sig}{k\vsig^2}$ into $\bar{\vsig}^i(\qq,\sig)=\vsig^i(\qq,\sig)-\frac{\qq}{3}$, $i=1,2,3$. Expression for $\frac{d\bar{\vsig}^i}{d\qq}$ can be obtained either by direct differentiation of the previously obtained expression for $\vsig^i(\qq,\sig)$ or subtracting $\frac{1}{3}$ from $\frac{d\vsig}{d\qq}=\frac{k\vsig^3}{\vsig^3+2\sig}$.\\
\noindent (b) is obvious.\\
\noindent To prove (c), recall, that $k<\frac{1}{3}$, and use formulas from (a).
\noindent To prove (d), recall, that $k<\frac{1}{3}$, and use the first formula from (a).
\noindent (e) immediately follows from (a) and (c).  \hfill $\Box$\\

\begin{lem}\label{l5}Solutions, $\vsig^1(0,\sig), \vsig^2(0,\sig), \vsig^3(0,\sig)$, of the equation $G(\vsig,0,\sig)=\vsig^3+\vsig^2-\sig=0$, $0<\sig\leq \frac{4}{27}$, enjoy the following properties:

\begin{description}
  \item[(a)] If $\sig=\frac{4}{27}$ the solutions are $\vsig^1(0,\frac{4}{27})=\frac{1}{3}$, $\vsig^2(0,\frac{4}{27})=\vsig^3(0,\frac{4}{27})=-\frac{2}{3}$
  \item[(b)] If $0<\sig_1<\sig_2\leq 0$, then $$0<\vsig^1(0,\sig_1)<\vsig^1(0,\sig_2)\leq\frac{1}{3}$$ and
  $$-1< \vsig^3(0,\sig_1)<\vsig^3(0,\sig_2)\leq -\frac{2}{3} \leq \vsig^2(0,\sig_2)<\vsig^2(0,\sig_1)<0$$
  \item[(c)] $\vsig^1(0,\sig)+\vsig^2(0,\sig)<0$
\end{description}

\end{lem}
\noindent {\bfseries Proof}. (a) can be checked directly. \\
\noindent To prove (b), one can either apply the implicit function theorem to $H(\vsig,\sig)=G(\vsig,0,\sig)=0$.
Or, less formally, draw the graph of $y=\vsig^3+\vsig^2-\frac{4}{27}$ and observe what happens to its roots when the graph is shifted upward until it becomes $y=\vsig^3+\vsig^2$.\\
\noindent (c) Obviously, $\vsig^1(0,\sig)+\vsig^2(0,\sig)+\vsig^3(0,\sig)=-1$. So, $\vsig^1(0,\sig)+\vsig^2(0,\sig)=-1 - \vsig^3(0,\sig)<0$, since $\vsig^3(0,\sig)>-1$.

\subsection{Solutions of the St. Venant-Kirchhoff Material}
We are now ready to present our main results.

\begin{thm}\label{p1}For any given force field $\bff:\Oo \rightarrow \real^d$ and  the
 surface traction $\bt : \Gt \rightarrow \real^d$ such that
the   statically admissible stress  $\btau \in \calT_a$ has no zero eigenvalues almost ever where  in  $\Oo$,
 the canonical dual problem $(\calP^d)$ 
 has a unique positive critical solution $\bT \in \calS^+_a$.
\end{thm}
\noindent {\bfseries Proof}. We need to prove that for arbitrarily given  $\sig_1,\sig_2,\sig_3>0$,
the system of equations (\ref{mainsystem1}) has a unique positive solution $(\vsig_1,\vsig_2,\vsig_3)$, such that all $\vsig_i>0$, $i=1,2,3$.
 From Lemma \ref{l4}(b), it follows that
$$F^{1,1,1}(0,\sig_1,\sig_2,\sig_3)>0 .$$
From Lemma \ref{l4}(c), it follows that for some $\qq_1>0$, large enough,
$$F^{1,1,1}(\qq_1,\sig_1,\sig_2,\sig_3)<0 .$$
Therefore, since $F^{1,1,1}$ is continuous and monotonically decreasing in $\qq$ (Lemma \ref{l4}(e)), there exists a unique $\qq_0$, $0< \qq_0<\qq_1$, such that
$$F^{1,1,1}(\qq_0,\sig_1,\sig_2,\sig_3)=0 .$$
i.e. $$\vsig^1(\qq_0,\sig_1)+\vsig^1(\qq_0,\sig_2)+\vsig^1(\qq_0,\sig_3)=\qq_0 . $$
So, from Lemma \ref{l1} it follows that $\vsig^1(\qq_0,\sig_1)$, $\vsig^1(\qq_0,\sig_2)$, $\vsig^1(\qq_0,\sig_3)$ form a positive solution of (\ref{mainsystem1}), which are eigenvalues of the second Piola-Kirchhoff stress $\bT$.
Therefore, Problem $(\calP^d)$ has a unique global maximizer $\bT \in \calS^+_a$.  \hfill $\Box$

\begin{thm}\label{p2}For any given force field $\bff:\Oo \rightarrow \real^d$ and  the
 surface traction $\bt : \Gt \rightarrow \real^d$ such that the eigenvalues of the
 statically admissible stress tensor function
 $\btau \in \calT_a$  satisfy
 $0<\sig_1,\sig_2,\sig_3<\frac{4}{27}$,
 the total complementary energy $\Pi^d_\btau (\bT) $
  has eight  negative solutions $\bT_k \in \calS^-_a$, $k=1, \dots, 8$.
\end{thm}
\noindent {\bfseries Proof}.
We need to prove that for arbitrarily given  $0<\sig_1,\sig_2,\sig_3<\frac{4}{27}$, the  system of equations (\ref{mainsystem1}) has 8 solutions
 $(\vsig_1,\vsig_2,\vsig_3)$, such that all $\vsig_i<0$, $i=1,2,3$.
From Corollary \ref{cor2l2} it follows that each of the equations $G(\vsig,0,\sig_i)$, has two negative solutions, $\vsig^2(0,\sig_i)>\vsig^3(0,\sig_i)$, $i=1,2,3$.
From Lemma \ref{l4}(b), it follows that for $i,j,k=2,3$
$$F^{i,j,k}(0,\sig_1,\sig_2,\sig_3)<0 .$$

\noindent From Lemma \ref{l4}(d) it follows that there exists $\qq_1<0$ such that
\[
F^{i,j,k}(\qq_1,\sig_1,\sig_2,\sig_3)>0 .
\]
Therefore, since $F^{i,j,k}$ is continuous in $\qq$ (Lemma \ref{l4}(e)), there exists $\qq_0$, $0> \qq_0>\qq_1$, such that $$F^{i,j,k}(\qq_0,\sig_1,\sig_2,\sig_3)=0 , $$
i.e.
\[
\vsig^i(\qq_0,\sig_1)+\vsig^j(\qq_0,\sig_2)+\vsig^k(\qq_0,\sig_3)=\qq_0 .
\]
So, from Lemma \ref{l1} it follows that $\vsig^i(\qq_0,\sig_1)$, $\vsig^j(\qq_0,\sig_2)$, $\vsig^k(\qq_0,\sig_3)$ form a negative solution of (\ref{mainsystem1}).
Since, each of $i,j,k$ can be chosen independently from the set $\{2,3\}$, we have total
8 different negative solutions.  \hfill $\Box$

\begin{thm}\label{p2}
For any given force field $\bff:\Oo \rightarrow \real^d$ and  the
 surface traction $\bt : \Gt \rightarrow \real^d$ such that the eigenvalues of the
 statically admissible stress tensor function
 $\btau \in \calT_a$  satisfy
 $0<\sig_1,\sig_2,\sig_3<\frac{4}{27}$,
 the total complementary energy $\Pi^d_\btau (\bT) $
 has at least 15 mixed stationary points, i.e., some  eigenvalues  $\vsig_i$, $i=1,2,3$,   of $\bT$     are positive, some  are negative.
\end{thm}
\noindent {\bfseries Proof}. Each of the equations $G(\vsig,0,\sig_i)$, has one positive and two negative solutions: $\vsig^1,\vsig^2,\vsig^3$.
 \noindent Applying Lemma  \ref{l5} it is easy to check that

 \begin{description}
  \item[(1)]  for  $i,j=2,3$, $$F^{1,i,j}(0,\sig_1,\sig_2,\sig_3)<0, F^{i,1,j}(0,\sig_1,\sig_2,\sig_3)<0$$
  \item[(2)]  $F^{2,3,1}(0,\sig_1,\sig_2,\sig_3)<0$, $F^{3,2,1}(0,\sig_1,\sig_2,\sig_3)<0$, $F^{3,3,1}(0,\sig_1,\sig_2,\sig_3)<0$.
  \item[(3)] $F^{1,1,2}(0,\sig_1,\sig_2,\sig_3)<0$, $F^{1,1,3}(0,\sig_1,\sig_2,\sig_3)<0$, $F^{1,3,1}(0,\sig_1,\sig_2,\sig_3)<0$, \\
  $F^{3,1,1}(0,\sig_1,\sig_2,\sig_3)<0$.
\end{description}

 \noindent For each of these 15 combinations, $F^{a,b,c}$, there exists $q_1<0$ such that  $F^{a,b,c}(q_1,\sig_1,\sig_2,\sig_3)>0$. \\
Therefore, since $F^{a,b,c}$ is continuous in $q$ (Lemma \ref{l4}(e)), there exists $q_0$, $0> q_0>q_1$, such that $$F^{a,b,c}(q_0,l_1,l_2,l_3)=0$$
that is
 $$\vsig^a(q_0,\sig_1)+\vsig^b(q_0,\sig_2)+\vsig^c(q_0,\sig_3)=q_0$$
So, from Lemma \ref{l1} it follows that $\vsig^a(q_0,\sig_1)$, $\vsig^b(q_0,\sig_2)$, $\vsig^c(q_0,\sig_3)$ form a mixed solution of (\ref{mainsystem1}).\\
Obviously, these 15 combinations result in different mixed stationary points of $\Pi^d_{\btau}(\bT)$. \hfill $\Box$\\

\section{Conclusions} \label{finish}
We have illustrated that  by using the canonical duality theory,
the nonconvex minimal potential problem $(\calP)$ is canonically dual to a
 concave maximization problem in a convex stress space $\calS_a^+$,
 which can be solved  by  well-developed numerical methods.
By the   pure complementary energy principle, the
 general nonlinear partial differential equation in nonlinear elasticity
is actually equivalent to  an algebraic (tensor) equation, which can be solved for certain materials to
obtain all possible stress solutions. Both global and local extremal solutions can be identified by the triality theory,
while the Legendre-Hadamard condition is only necessary for local minimizers.
Our results shows that for St. Venant-Kirchhoff material,
the nonlinear boundary value problem could have
 24 solutions at each material point, but only one global minimizer if the statically admissible stress $\btau \neq 0$.
It is important to have a detailed study on these solutions  in the future. \\

\noindent {\bf Acknowledgements}\\
The research of the first author was supported by
 the US Air Force Office of Scientific Research under the grant AFOSR FA9550-10-1-0487.
 Results presented in Section 3 were discussed with Professor Ray Ogden from University of Glasgow.


\begin{thebibliography}{99}

\bibitem {Abey1980}
Abeyaratne, R.~C. 1980 Discontinuous deformation gradients in plane
finite elastostatics of incompressible materials. \emph{J. Elasticity} \textbf{10}, 255--293.

\bibitem {Abe}
Abeyaratne, R.~C. 1981 Discontinuous deformation gradients in the
finite twisting of an incompressible elastic tube. \emph{J. Elasticity} \textbf{11}, 43--80.

\bibitem{arthus} Arthurs, A.M. (1980).
{\em  Complementary Variational Principles,} Clarendon Press,
Oxford.

\bibitem{cai-gao-qin}
Cai, K., Gao, DY, Qin, QH (2014). Post-buckling solutions of hyper-elastic beam by canonical dual finite element method,
{\em Mathematics and Mechanics of Solids}.   19(6) 659-671

\bibitem{ciarlet} Ciarlet, P.G. (1988). {\em Mathematical Elasticity. Volume I: Three-Dimensional Elasticity}.
North-Holland.

\bibitem {gao-zamp92} Gao, D.~Y. 1992 Global extremum criteria for nonlinear elasticity.
\emph{J. Appl. Math. Physics ( ZAMP)} \textbf{43}, 924--937.

\bibitem {gao-amr97} Gao, D.~Y. 1997 Dual extremum principles in finite deformation theory with applications to
post-buckling analysis of extended nonlinear beam theory.
\emph{Appl. Mech. Rev.} \textbf{50}, S64--S71.

\bibitem{gao-ima98} Gao, D.~Y. 1998 Duality, triality and complementary extremum  principles in nonconvex parametric variational
 problems with applications.
 \emph{IMAJ. Appl. Math.} \textbf{61}, 199--235.

\bibitem {gao-mrc99} Gao, D.~Y. 1999a Pure complementary energy principle and triality theory in finite elasticity.
\emph{Mech. Res. Comm.} \textbf{26}, 31--37.

 \bibitem {gao-wiley99}   Gao, D.~Y.  1999b  \emph{Duality-Mathematics}.
Wiley Encyclopedia of Electronical and Electronical Engineering
\textbf{6},  68--77.

\bibitem {gao-mecc99}   Gao, D.~Y. 1999c  General analytic solutions and
  complementary variational principles for large deformation
nonsmooth mechanics. \emph{Meccanica}  \textbf{34}, 169--198.

\bibitem {gao-dual00}  Gao, D.~Y. 2000a  \emph{Duality Principles in Nonconvex
Systems: Theory, Methods and Applications}. Kluwer Academic
Publishers,  Dordrecht /Boston /London, xviii + 454pp.

\bibitem {gao-na00}   Gao, D.~Y.  2000b  Analytic solution and triality
 theory for nonconvex and nonsmooth variational problems with applications.  \emph{Nonlinear Analysis}
 \textbf{42}, 1161--1193.

 \bibitem {gao-jogo00} Gao, D.~Y.  2000c  Canonical dual transformation method and
generalized triality theory in nonsmooth global optimization \emph{J. Global Optimization} \textbf{17}, 127--160.

 \bibitem {gao-amma03} Gao, D.~Y.  2003a
 Nonconvex semi-linear problems and canonical dual solutions.
 \emph{Advances in Mechanics and Mathematics}, Vol. II,
 D.~Y. Gao \& R.~W. Ogden (eds), Kluwer Academic Publishers,  pp. 261--312.

 \bibitem {gao-optm03} Gao, D.~Y.  2003b  Perfect duality theory and complete set of
 solutions to a class of global optimization.
 \emph{Optimization} \textbf{52}, 467--493.

\bibitem {gao-mms04}Gao, D.~Y.  2004  Complementary variational principle,
algorithm, and complete solutions to phase transitions in solids governed by Landau-Ginzburg  equation.
\emph{Math. Mech. Solids} \textbf{9}, 285--305.

\bibitem{gao-cace09} Gao, D.Y.  2009  Canonical duality theory: unified understanding and generalized solutions for
global optimization. {\em Comput. \& Chem. Eng.} 33,
1964-1972.


\bibitem{gao-ogden-qjmam08}
Gao, D.~Y. \& Ogden, R.~W.  2008.  Multiple solutions to non-convex variational problems with implications for phase transitions and numerical computation.
\emph{Q.  J. Mech. Appl. Math.} \textbf{61}, 497--522.


\bibitem{gao-ogden-zamp} Gao, D.Y. and Ogden, R.W.  2008.
Closed-form solutions, extremality and nonsmoothness criteria in a large deformation elasticity problem,
{\em ZAMP,} 59:498 - 517.

\bibitem {gao-ruan-jogo08} Gao, D.~Y. \& Ruan, N. 2010  Solutions to quadratic minimization problems with box and  integer
constraints. \emph{J. Global Optim.} \textbf{47}, 463--484.

\bibitem{gao-review14}  Gao, D.~Y.,   Ruan,  N. ~and Latorre, V. 2015.
Canonical duality-triality: Bridge between nonconvex analysis/mechanics and global optimization,
{\em Math. Mech. Solids}.







 \bibitem {gao-strang89a} Gao, D.~Y. \& Strang, G. 1989
Geometric nonlinearity: potential energy, complementary energy, and the gap function.
\emph{Q. Appl. Math.} \textbf{47}, 487--504.


\bibitem{gao-yu} Gao, D.Y., Yu, H.F. (2008).
Multi-scale modelling and canonical dual finite element method in
phase transitions of solids. \emph{ Int. J. Solids Struct.} 45, 3660-3673.

\bibitem {hell-14}  Hellinger, E.  1914
Die allgemeine Ans\"{a}tze der Mechanik der Kontinua.
Encyklop\"{a}die der Mathematischen Wissenschaften IV, 4,  602--694.

\bibitem {jian1998}
Jiang, X. \& Ogden, R.~W. 1998 On azimuthal shear of a circular
cylindrical tube of compressible elastic material. \emph{Q. J.
Mech. Appl. Math.} \textbf{51}, 143--158.

\bibitem {Kass}
Kassianidis, F., Merodio, J.,  Ogden, R.~W. \& Pence, T.~J. 2008 Azimuthal shear of a
transversely isotropic elastic solid. \emph{Math. Mech. Solids } \textbf{13}, 690--724.

\bibitem {koiter76}  Koiter, W.~T.  1976
 On the complementary energy theorem in nonlinear elasticity theory. In \emph{Trends in Appl. of Pure Math. to Mech.},   G. Fichera (ed.), Pitman.

\bibitem {lee-shie80} Lee, S.~J. \& Shield, R.~T.  1980a
Variational principles in finite elastics. \emph{J. Appl. Math. Physics (ZAMP)} \textbf{ 31}, 437--453.

\bibitem {lee-shie80b} Lee, S.~J. \& Shield, R.~T.  1980
Applications of variational principles in finite elasticity.
\emph{J. Appl. Math. Physics (ZAMP)} \textbf{ 31},  454--472.

\bibitem{levi-65} Levinson, M.  1965
The complementary energy theorem in finite  elasticity.
\emph{J. Appl. Mech.}  \textbf{87},  826--828.

\bibitem {li-cupta} Li, S.~F. \& Gupta, A. 2006 On dual configuration forces.
\emph{J. Elasticity} \textbf{84}, 13--31.


\bibitem{li-zhou-gao} Li, C., Zhou, X., and Gao DY (2014).
Stable trajectory of logistic map, {\em Nonlinear Dynamics},
DOI 10.1007/s11071-014-1433-y

\bibitem{nobel} Noble, B. and Sewell, M.J.  (1972).
On dual extremum principles in applied mathematics,
{\em J. Inst. Math. Appl.,} 9, 123-193.

 \bibitem {oden-redd83}  Oden, J.~T. \& Reddy, J.~N. 1983
\emph{Variational Methods in Theoretical Mechanics}. Springer-Verlag.

\bibitem {ogden75} Ogden, R.~W.  1975
A note on variational theorems in non-linear elastostatics.
\emph{Math. Proc. Camb. Phil. Soc.} \textbf{77}, 609--615.

 \bibitem {ogden77} Ogden, R.~W.  1977
Inequalities associated with the inversion of elastic  stress-deformation relations and their implications.  \emph{Math. Proc. Camb.  Phil. Soc.} \textbf{81},  313--324.

\bibitem {reiss53} Reissner, E.  1953
On a variational theorem for finite elastic deformations.
\emph{J. Math. Phys.} \textbf{32}, 129--135.

\bibitem {rivl1949}
Rivlin, R.~S.  1949 Large elastic deformations of isotropic materials VI. Further results in the theory of torsion, shear and flexure.
\emph{Phil. Trans. R. Soc. Lond. A} \textbf{242}, 173--195.

\bibitem {gao-ruan-pardalos}  Ruan, N. and Gao, D.Y. (2014). Global optimal solutions to a general
sensor network localization problem,   {\em Performence Evaluations},  75-76: 1-16.


\bibitem{ruan-gao-ima} Ruan, N. and Gao, D.Y.(2014).  Canonical duality approach for nonlinear dynamical systems,
{\em IMA J. Appl. Math.},  79: 313-325.

\bibitem {ruan-gao-jiao-coap08} Ruan, N., Gao, D.~Y. \& Jiao, Y. 2010 Canonical dual least square method for solving general nonlinear
 systems of equations. \emph{Comput. Optim. Appl.}
 \textbf{47}, 335--347.

\bibitem{hugo-gao}  Santos, H.A.F.A. and Gao D.Y. (2011).
 Canonical dual finite element method for solving post-buckling problems of a large deformation elastic beam, {\em Int. J. Nonlinear Mechanics, } 47: 240 - 247. doi:10.1016/j.ijnonlinmec.2011.05.012

\bibitem{Tabarrok-Rimrott} Tabarrok, B. and   Rimrott, F.P. (1994).
{\em Variational Methods and Complementary Formulations in Dynamics}, Springer, 366pp.

\bibitem{Truesdell-Noll}
Truesdell C. and Noll W.:
{\em  The Non-Linear Field Theories of Mechanics.} Springer Verlag, third
edition, 2004.

\bibitem {veub72} Veubeke, B.~F.  1972
A new variational principle for finite elastic displacements.
\emph{ Int. J. Eng. Sci.} \textbf{10}, 745--763.

\bibitem{zgy}Zhang J., Gao, D.Y. and Yearwood, J. (2011).
 A novel canonical dual computational approach for prion AGAAAAGA amyloid fibril molecular modeling. {\em Journal of Theoretical Biology}, 284,  149-157  (2011). doi:10.1016/j.jtbi.2011.06.024

\end{thebibliography}
\end{document}